
\input harvmac

\def\ch{{\rm ch}}   \def\td{{\rm td}}
\def\e{{\rm e}}     \def\p{{\rm p}}
\def\im{{\rm Im\,}} \def\ker{{\rm Ker\,}}
\def\s{{\rm s}}     \def\T{{\rm T}}
\def\D{{\rm D}}     \def\L{{\rm L}}
\def\half{\hbox{$1\over2$}}
\def\inbar{\,\vrule height1.5ex width.4pt depth0pt}
\def\C{\relax{\hbox{$\inbar\kern-.3em{\rm C}$}}}     
\def\Box{{\hbox{$\sqcup$}\llap{\hbox{$\sqcap$}}}}

\Title{DAMTP R92/42}
{\vbox{\centerline{Topological Conformal Gravity}
 \vskip4pt\centerline{in Four Dimensions}}}

\centerline{Malcolm J. Perry~ and~ Edward Teo}
\bigskip\centerline{Department of Applied Mathematics
 and Theoretical Physics}
\centerline{University of Cambridge}\centerline{Silver Street}
\centerline{Cambridge CB3 9EW}\centerline{England}

\vskip .5in
\centerline{\bf Abstract}

\noindent
In this paper, we present a new formulation of topological
conformal gravity in four dimensions. Such a theory was first
considered by Witten as a possible gravitational counterpart of
topological Yang--Mills theory, but several problems left it
incomplete. The key in our approach is to realise a theory which
describes deformations of conformally self-dual gravitational
instantons.  We first identify the appropriate elliptic complex
which does precisely this. By applying the Atiyah--Singer index
theorem, we calculate the number of independent deformations of
a given gravitational instanton which preserve its self-duality.
We then quantise topological conformal gravity by BRST gauge-fixing,
and discover how the quantum theory is naturally described by the
above complex. Indeed, it is a process which closely parallels
that of the Yang--Mills theory, and we show how the partition
function generates an uncanny gravitational analogue of the
first Donaldson invariant.

\Date{November 92}

\newsec{Introduction}

It is somewhat ironic to note that in recent years, physicists
have applied the Feynman path integral formalism in quantum field
theory to derive various non-trivial differential-topological
results in mathematics. Such theories are generally known as
topological quantum field theories (a review may be found in
ref.~\ref\rBBRT{D.~Birmingham, M.~Blau, M.~Rakowski and
G.~Thompson, Phys.~Rep. 209 (1991) 129}), and have largely arisen
from the work of Witten.

In ref.~\ref\rTQFT{E.~Witten, Commun.~Math.~Phys. 117 (1988) 353},
Witten introduced a type of supersymmetric Yang--Mills theory
whose partition function generates the Donaldson polynomial
invariants of smooth four-manifolds \ref\rDonaldson{S.K.~Donaldson
and P.B.~Kronheimer, The Geometry of Four-Manifolds (Clarendon
Press, Oxford, 1990)}. The essential point of this model is the
introduction of ghost fields to cancel any local dynamical
degrees of freedom. This produces a quantum field theory which
does not depend on any local structure like the metric, but whose
observables are manifold invariants. Such invariants characterise
the global properties of the manifold in question, which may be
its topology, or its underlying smooth or differential structure
as in the case of the Donaldson polynomials. The analogue of this
theory in two dimensions is a certain non-linear, supersymmetric
sigma model \ref\rTSM{E.~Witten, Commun.~Math.~Phys. 118 (1988)
411}, which Witten has shown to give rise to invariants of
symplectic geometry due to Gromov. In a third paper \ref\rTKT{
E.~Witten, Commun.~Math.~Phys. 121 (1988) 351}, Witten also
showed how the three-dimensional Chern--Simons action produces
the Jones polynomial of knot theory.

While the techniques used in quantum field theory will probably
not stand up to a completely mathematically rigorous treatment,
mathematicians may still profitably utilise these physical
constructions to gain insight into both presently known and
new manifold invariants. A prime example of this is the use
of Yang--Mills instantons by Donaldson \rDonaldson\ to construct
his polynomial invariants. Conversely, topological quantum field
theories provide fertile ground for physicists to deepen the
understanding of the (in)formal manipulations of Feynman path
integrals, especially when using them to ``derive'' rigorous
results in mathematics like the Donaldson and Jones polynomials.

It is with this philosophy in mind that we write this paper. Back
in four dimensions, Witten \ref\rTG{E.~Witten, Phys.~Lett. B206
(1988) 601} has considered conformal gravity as a candidate for
constructing a gravitational counterpart of the Donaldson
polynomials. However, the model he constructed was unsatisfactory
in a few ways, and he thus left open the question of whether
his theory led to any invariants of four-manifolds. In this
paper, we will attempt to rebuild this model from a new point
of view. Throughout, we will draw upon the many remarkable
similarities between topological Yang--Mills theory and its
conformal gravity analogue. Indeed, it is surprising that while
self-dual Yang--Mills instantons over four-manifolds play a very
important r\^ole in the recent study of these manifolds, self-dual
gravitational instantons {\it intrinsic} to four-manifolds have
yet to make an impact on this area. We will thus propose
suitable gravitational analogues of the Donaldson polynomials
as possibly new invariants of four-manifolds.

Let us now give a lightning review of several important points of
topological Yang--Mills theory \refs{\rTQFT,\rBBRT}\ that will be
relevant to our discussion below. Consider the classical action
\eqn\eClassicalTYM{\int_M \hbox{d}^4x\ \tr\{F_{ab}\ast F^{ab}\}\ ,}
where $F_{ab}$ is the usual Yang--Mills field strength of the
gauge potential $A_a$. The trace is over the gauge-group indices,
and $M$ is a compact four-manifold. $\ast F_{ab}={1\over2}
\epsilon_{ab}{}^{cd}F_{cd}$ is the dual of $F_{ab}$. This action
is invariant under arbitrary variations of the gauge field $A_a$.
In order to quantise it using the Feynman path integral
formalism, the topological symmetry of the action has to be
gauge-fixed. This is done using the BRST quantisation method,
and it introduces three quantum ghost fields to completely break
the symmetry. With an appropriate choice of gauge \ref\rBS{
L.~Baulieu and I.M.~Singer, Nucl.~Phys.~B (Proc.~Suppl.) 5B
(1988) 12}, the final gauge-fixed action turns out to be the
ordinary Yang--Mills functional
\eqn\eClassicalYM{\int_M \hbox{d}^4x\ \tr\{F_{ab}F^{ab}\}\ ,}
plus extra ghost interaction terms which cancel out all local
degrees of freedom in \eClassicalYM.

Now, the classical action \eClassicalYM\ is minimised by field
strengths which satisfy either the self-duality condition
$F_{ab}=\ast F_{ab}$, or the anti-self-duality condition
$F_{ab}=-\ast F_{ab}$. Their corresponding gauge potentials
are known as Yang--Mills instantons. Thus, in evaluating the
partition function and quantum observables of topological
Yang--Mills theory, we have to consider quantum fluctuations
around these classical instanton solutions. In other words, we
need a theory describing self-dual, infinitesimal {\it
deformations} around a given Yang--Mills instanton.

Such a theory was first studied by Atiyah, Hitchin and Singer \ref
\rAHS{M.F.~Atiyah, N.J.~Hitchin and I.M.~Singer, Proc.~R.~Soc.~Lond.
A362 (1978) 425} in the more general context of the Yang--Mills
moduli space, that is the space of all gauge-inequivalent
instantons.  This moduli space was a vital tool used by Donaldson
in discovering his manifold invariants \rDonaldson. For our
present purposes, the most important property of it that we need
to know is its dimension. Observe that the value of the dimension
is precisely the number of independent non-trivial deformations
that can be made around an instanton, which preserve its
self-duality. (Physicists usually refer to this number as the
number of free parameters of the instanton solution.) It
depends only on the topological properties of the manifold
and fibre bundle in question.

The way to derive the value of this dimension is to introduce
a sequence of mappings between vector bundles, known as an
{\it elliptic complex}, which describes deformations of
instantons \rAHS. (This has also been done by physicists in
ref.~\ref\rBCGW{C.W.~Bernard, N.H.~Christ, A.H.~Guth and
E.J.~Weinberg, Phys.~Rev. D16 (1977) 2967}.) It is closely
related to the de~Rham complex of forms. By invoking the
Atiyah--Singer index theorem \ref\rShan{P.~Shanahan, The
Atiyah--Singer Index Theorem: An Introduction (Springer--Verlag,
Berlin, 1978)}, one can then calculate the so-called index of
this elliptic complex. The value of this index turns out to be
the dimension of the moduli space.

The nature of the partition function of topological Yang--Mills
theory depends crucially on the dimension of the Yang--Mills
moduli space \rTQFT. As it also turns out, the three gauge-fixing
ghost fields of the quantum action belong to each of the three
vector bundles of the elliptic complex. This is not too surprising
by remarks we made earlier, that these ghost fields characterise
quantum deformations about the classical instanton solutions of
\eClassicalYM. Thus, the importance of the mathematical theory
of instanton deformations to topological Yang--Mills theory
simply cannot be overlooked.

In this paper, we will ask ourselves whether we can proceed
analogously to construct a theory of topological gravity. In
ref.~\rTG, Witten proposed conformal gravity as the appropriate
counterpart to the Yang--Mills theory. This theory has, as the
classical action,
\eqn\eClassicalCG{\int_M \hbox{d}^4x\ \sqrt g\ C_{abcd}C^{abcd}\ ,}
where $C_{abcd}$ is the Weyl tensor of $M$. This is the
gravitational analogue of \eClassicalYM. It is minimised by
manifolds possessing either a self-dual or anti-self-dual Weyl
tensor, and such manifolds are known as conformally (anti-)
self-dual gravitational instantons \ref\rSHP{A.~Strominger,
G.T.~Horowitz, M.J.~Perry, Nucl.~Phys. B238 (1984) 653}.

The corresponding topological action for conformal gravity is
\eqn\eClassicalTG{\int_M \hbox{d}^4x\ \sqrt g\ C_{abcd}
\ast C^{abcd}\ ,}
clearly the analogue of \eClassicalTYM. We will construct its
quantum theory by BRST gauge-fixing, and this introduces three
ghost fields. Again with an appropriate choice of gauge, the total
quantum action turns out to be the sum of \eClassicalCG\ and a
set of ghost interaction terms, which cancel out all local
degrees of freedom in \eClassicalCG.

It is clear that we now need {\it a deformation theory of
conformally self-dual gravitational instantons}, to characterise
the quantum effects of the ghost fields. We write down an
elliptic complex which realises this. Actually, this complex was
considered by I.M.~Singer in unpublished work many years ago
\ref\rSinger{I.M.~Singer, unpublished (1977)}, but has since
remained relatively unknown in both the mathematics and physics
literature\foot{However, very recently, there emerged a rigorous
mathematical study into such conformally self-dual deformations
of gravitational instantons \ref\rKK{A.D.~King and D.~Kotschick,
The Deformation Theory of Anti-Self-Dual Conformal Structures, to
be published in Math.~Ann.}, in an effort to discover invariants
of four-manifolds just as Donaldson used Yang--Mills instantons
to construct his invariants.}. We will refer to this elliptic
complex as the gravitational instanton deformation complex.
By applying the Atiyah--Singer index theorem to this elliptic
complex, we derive an expression for the number of independent
deformations of a given manifold that preserve its self-duality.
This number $\varpi$ is essentially a topological quantity, which
will be important in the quantum theory of topological gravity.
Under certain conditions, one could then integrate these
infinitesimal deformations to obtain a local moduli space of
gravitational instanton metrics, whose dimension is $\varpi$.

Since Witten's original work on topological gravity \rTG,
there have been several papers written on such theories in
four dimensions. In fact, it was very quickly demonstrated
\nref\rLP{J.M.F.~Labastida and M.~Pernici, Phys.~Lett. B213
(1988) 319}\nref\rBMS{R.~Brooks, D.~Montano and J.~Sonnenschein,
Phys.~Lett. B214 (1988) 91}\refs{\rLP,\rBMS}\ how Witten's
quantum topological gravity theory arises from the BRST
gauge-fixing of the classical action \eClassicalTG. However,
as in the original paper itself, the significance of the
gravitational instanton deformation complex in this theory was
not appreciated. Consequently, topological gravity in four
dimensions became a nightmare of long spinor expressions and
complicated differential operators, and explicit invariants were
not found. What we have succeeded in doing, in this paper, is to
simplify the whole process of gauge-fixing topological gravity,
by rewriting everything in terms of the two differential operators
occurring in the deformation complex. This results in a procedure
that is no more complicated than gauge-fixing topological
Yang--Mills theory, and its interpretation becomes much clearer.
In particular, we are able to demonstrate how to calculate a
simple ``manifold invariant'', which is just the value of the
partition function itself, in the case when the gravitational
moduli space is discrete. It turns out to be the number of points
in this moduli space, counted with signs. This is strikingly
similar to the first Donaldson invariant in the Yang--Mills case
\rDonaldson.

\nref\rTorre{C.G.~Torre, Phys.~Rev. D41 (1990) 3620; Phys.~Lett.
B252 (1990) 242; A Deformation Theory of Self-Dual Einstein
Spaces, preprint (Sept., 1991)}

We perhaps should also mention that there has been some effort by
Torre \rTorre\ to construct a topological field theory based on
the deformation theory of gravitational instantons. This is in
much the same spirit as this paper, and indeed he has constructed
an elliptic complex to study these deformations. But while his
use of Ashtekar variables is a very good idea, he did not apply
the Atiyah--Singer index theorem to calculate the index of his
complex, and so was not led to a complete theory.

\nref\rMyers{R.~Myers and V.~Periwal, Nucl.~Phys. B361 (1991) 290}
\nref\rJap{A.~Nakamichi, A.~Sugamoto and I.~Oda, Phys.~Rev. D44
(1991) 3835\semi
I.~Oda and A.~Sugamoto, Phys.~Lett. B266 (1991) 280}

Several other groups \refs{\rMyers,\rJap}\ have also in fact
studied four-dimensional topological gravity, not constructed
from \eClassicalTG, but from topological combinations of the
Riemann tensor. Hence their models are not conformally invariant
like ours.

Before beginning in detail, let us clarify the spinor notation
used in this paper. The analogue of the Lorentz group on a manifold
with positive-definite signature is SO(4). Its two-fold
covering group is Spin(4), which has the decomposition Spin(4)
$\simeq$ SU(2) $\times$ SU(2). So at least locally, we can
introduce two vector bundles $\Omega_+$ and $\Omega_-$ of spinor
fields with positive and negative chirality. They are
also known as the bundles of anti-self-dual and self-dual
spinors respectively \rAHS.

General spin bundles may then be constructed:
\eqn\eBundleSplit{\Omega_{mn} = S^m\Omega_+ \otimes
S^n\Omega_-\ ,}
where $S^n\Omega_{\pm}$ is the $n$-fold symmetric product of
$\Omega_{\pm}$. A field $\Psi$ belonging to $\Omega_{mn}$ has
spin $(m,n)$, and transforms under the {\it irreducible}
representation $(m,n)$ of SU(2) $\times$ SU(2). In the notation
of two-component spinors, it has $m$ unprimed and $n$ primed
indices
\eqn\eMNfield{\Psi_{A_1...A_mA_1^\prime...A_n^\prime}\ ,}
which are completely symmetric in $A_1...A_m$ and $A_1^\prime
...A_n^\prime$. (By virtue of this symmetry, they are also
completely trace-free in $A_1...A_m$ and $A_1^\prime...
A_n^\prime$.) Such a field will be referred to as an
$(m,n)$-field. The dimension of the vector space $\Omega_{mn}$
is \nref\rCD{S.M.~Christensen and M.J.~Duff, Nucl.~Phys. B154
(1979) 301}\nref\rPR{R.~Penrose and W.~Rindler, Spinors and
Space-time: Vols.~I, II (Cambridge University Press, Cambridge,
1984)}\refs{\rCD,\rPR}
\eqn\eDimOmega{\dim\Omega_{mn}=(m+1)(n+1)\ .}

Each spinor index takes either value 0 or 1. In general, a
tensor $\Phi_{\alpha_1...\alpha_n}$ with $n$ indices can be
written as a spinor with $n$ unprimed and $n$ primed indices
$\Phi_{A_1...A_nA_1^\prime...A_n^\prime}$. We can further
decompose it into its irreducible components, which consist
of a spinor $\tilde\Phi_{A_1...A_nA_1^\prime...A_n^\prime}
\in\Omega_{nn}$, completely symmetric in $A_1...A_n$ and
$A_1^\prime...A_n^\prime$, plus other anti-symmetric terms.
These anti-symmetric terms consist of lower-order irreducible
spinors multiplied by an appropriate number of anti-symmetric
epsilon symbols $\epsilon_{AB}$ or $\epsilon_{A^\prime B^\prime}$.
Further details may be found in ref.~\rPR.

Let us quickly run through a few examples of the types of
spinors that will be used in this paper. A covector $\xi_a$
is written as $\xi_{A A^\prime}$ in spinor notation. For a less
trivial case, consider the symmetric two-index tensor $h_{ab}$.
It can be decomposed as
\eqn\eHDecomp{h_{ab}\equiv h_{ABA^\prime B^\prime} =
\tilde h_{ABA^\prime B^\prime} + \hbox{${1\over4}$}
h\epsilon_{AB}\epsilon_{A^\prime B^\prime}\ ,}
where the irreducible component $\tilde h_{ABA^\prime B^\prime}
\in\Omega_{22}$ describes the trace-free part of $h_{ab}$, and
$h\in\Omega_{00}$ is the trace part of $h_{ab}$. One can easily
check to see that the number of degrees of freedom correspond.
On the other hand, the anti-symmetric two-index tensor $F_{ab}$
may be decomposed spinorally as
\eqn\eFDecomp{F_{ab}\equiv F_{ABA^\prime B^\prime} = F_{AB}
\epsilon_{A^\prime B^\prime}+F_{A^\prime B^\prime}
\epsilon_{AB}\ .}
It turns out that the symmetric spinor $F_{AB}\in\Omega_{20}$
is anti-self-dual, while $F_{A^\prime B^\prime}\in\Omega_{02}$
is self-dual. This exemplifies the well-known fact that a 2-form
$F_{ab}$ may be written as a direct sum of its self-dual and
anti-self-dual parts.

{}From now on, all the spinors encountered in our study of
topological gravity will be irreducible ones. The one exception
is the metric tensor, which in spinor notation reads $g_{ab}
\equiv g_{ABA^\prime B^\prime}=\epsilon_{AB}\epsilon_{A^\prime
B^\prime}$. It does not belong to $\Omega_{22}$, although its
trace-free variation does. The spinors we will use are also
globally well-defined, even on a manifold which does not admit
a spin structure.

Our spinor notation generally follows that of ref.~\rPR, except
our Ricci spinor $\Phi_{ABA^\prime B^\prime}$ and Ricci scalar $R$
both have opposite signs to their notation. This is to conform to
what is more usual in the literature.

\newsec{Deformations of a conformally self-dual manifold}

On a four-dimensional Riemannian manifold, 2-forms are special
in that the space of 2-forms, $\Lambda^2$, splits into the direct
sum
\eqn\eTwoForms{\Lambda^2=\Lambda_+^2\oplus\Lambda_-^2\ ,}
of self-dual and anti-self-dual 2-forms respectively. The
Riemann curvature tensor is a self-adjoint transformation
${\cal R}:\Lambda^2\rightarrow\Lambda^2$ given by \nref\rEGH{
T.~Eguchi, P.B.~Gilkey and A.J.~Hanson, Phys.~Rep. 66 (1980)
213}\refs{\rAHS,\rEGH}
\eqn\eRiemann{{\cal R}(e^a\wedge e^b)=\half\sum_{c,d}R^{ab}{}
_{cd}e^c\wedge e^d\ ,}
where $\{e^a\}$ is some basis of 1-forms. Hence we can block
diagonalise the transformation ${\cal R}$ with respect to the
decomposition \eTwoForms
\eqn\eRiemDecomp{{\cal R}=\left[\matrix{X&Y\cr Y^\ast&Z}
\right],}
where $X:\Lambda_+^2\rightarrow\Lambda_+^2$, $Z:\Lambda_-^2
\rightarrow\Lambda_-^2$ and $Y:\Lambda_-^2\rightarrow\Lambda_+^2$.
In terms of its irreducible components, ${\cal R}$ may be
decomposed into the trace-free part of $X$, the trace-free part of
$Z$, the trace part, and $Y$. In more familiar notation, this
decomposition may be written as
\eqn\eRiemDecompI{R^{ab}{}_{cd}=W_{(+)}{}^{ab}{}_{cd} + W_{(-)}
{}^{ab}{}_{cd} + \hbox{${1\over6}$}\delta^{[a}{}_{[c}\delta^{b]}
{}_{d]}R + 2\delta^{[a}{}_{[c}\Phi^{b]}{}_{d]}\ .}
$R$ is the Ricci scalar, $\Phi_{ab}$ is the trace-free part of
the Ricci tensor, $W_{(+)}{}_{abcd}$ is the self-dual part of the
Weyl tensor, and $W_{(-)}{}_{abcd}$ its anti-self-dual part.
The complete Weyl tensor is
\eqn\eWeylTensor{C_{abcd}=W_{(+)}{}_{abcd}+W_{(-)}{}_{abcd}\ .}
It is self-dual:
\eqn\eConformSD{C_{abcd}=\ast C_{abcd}\equiv\half\epsilon_{ab}
{}^{ef}C_{efcd}\ ,}
if and only if $W_{(-)}{}_{abcd}=0$, and is anti-self-dual:
\eqn\eConformASD{C_{abcd}=-\ast C_{abcd}\equiv-\half\epsilon_{ab}
{}^{ef}C_{efcd}\ ,}
if and only if $W_{(+)}{}_{abcd}=0$. Note that if the Riemann
tensor is self-dual, then it follows from the irreducible
decomposition \eRiemDecompI\ that the Weyl tensor is also
self-dual, and that the full Ricci tensor vanishes. The converse
is however untrue---the self-duality of the Weyl tensor {\it does
not} imply the self-duality of the Riemann tensor. We will be
interested in the more general case of the Weyl tensor being
self-dual. A manifold with metric admitting such a property
is known as a conformally self-dual manifold, and some examples
will be discussed below.

I.M.~Singer has, in unpublished work \rSinger, derived an
expression for the number $\varpi$ of independent conformally
self-dual deformations of a given compact manifold $M$, that has
a metric which is conformally self-dual. $\varpi$ is the sum of
a topological quantity and two terms dependent on the metric on
$M$, given by \rEGH
\eqn\eDimMSp{\eqalign{\varpi &= \half(29|\tau|-15\chi) +
\hbox{(dimension of the conformal group of $M$)} \cr &\quad
+ \hbox{(correction for absence of vanishing theorem if $R\leq
0$)}\ .}}
$\varpi$ was derived by applying the Atiyah--Singer index theorem
to an appropriate elliptic complex. $\chi$ and $\tau$ are the
Euler characteristic and Hirzebruch signature of the manifold
respectively, both of which are topological invariants. They can
be written in terms of the irreducible components of the Riemann
tensor as
\eqn\eChi{\chi=\hbox{$1\over32\pi^2$}\int\hbox{d}^4x\ \sqrt g\
(C_{abcd}C^{abcd}-2\Phi_{ab}\Phi^{ab} + \hbox{$1\over6$}R^2)\ ,}
and
\eqn\eTau{\tau=\hbox{$1\over48\pi^2$}\int\hbox{d}^4x\ \sqrt g\
C_{abcd}\ast C^{abcd}\ .}
The results hold if $M$ has no boundary, and will be modified
by appropriate boundary terms otherwise. The above expression
for $\varpi$ also excludes diffeomorphisms and conformal
rescalings of the metric, which trivially leave the manifold
conformally self-dual.

In the rest of this section, we will present a low-key
derivation of \eDimMSp\ and apply it to a few examples.

\subsec{The gravitational instanton deformation complex}

Consider the following sequence of mappings, which we call the
gravitational instanton deformation complex:
\eqn\eSingerCplxI{0 \ {\buildrel \D_{-1}\over\longrightarrow} \
\left\{\matrix{\hbox{infinitesimal} \cr
\hbox{coordinate} \cr \hbox{transformations}}\right\}
\ {\buildrel \D_0\over\longrightarrow} \
\left\{\matrix{\hbox{trace-free} \cr
\hbox{metric} \cr \hbox{variations}}\right\}
\ {\buildrel \D_1\over\longrightarrow} \
\left\{\matrix{\hbox{anti-self-dual} \cr
\hbox{part of} \cr\hbox{Weyl tensor}}\right\}
\ {\buildrel \D_2\over\longrightarrow} \ 0\ .}
This sequence is only defined on a manifold whose metric has a
{\it self-dual} Weyl tensor, although the theory of this section
may be carried through analogously for the anti-self-dual case
via a change in orientation.

$\D_{-1}$ is an operator which takes $0$ to the zero vector field.

$\D_0$ is defined by
\eqn\eDO{[\D_0\xi]_{ab}\equiv\nabla_a\xi_b + \nabla_b\xi_a -
\half g_{ab}\nabla_c\xi^c,}
where $\xi^a$ is a vector field generating an infinitesimal
coordinate transformation of $M$ via $x^a\rightarrow x^a+\xi^a$.
For this transformation to be a conformal one, it is necessary
and sufficient that $\xi_a$ satisfies the conformal Killing
vector equation
\eqn\eConformKV{\nabla_a\xi_b + \nabla_b\xi_a -
\half g_{ab}\nabla_c\xi^c=0\ .}
Hence the kernel of $\D_0$ is simply the conformal group of $M$.
The image of $\D_0$ is the trace-free part of the metric variation
under a change of coordinates:
\eqn\eLongitudinalP{\delta g_{ab} = \nabla_a\xi_b +
\nabla_b\xi_a - \half g_{ab}\nabla_c\xi^c.}
The trace component, which corresponds to conformal rescalings of
the metric under which the Weyl tensor is invariant, has been
projected out from the deformation complex.

We set $\D_1\equiv p_-\L$, where $p_-={1\over2}(1-\ast)$ is the
anti-self-dual projection operator. $\L$ is the operator defined
by
\eqn\eDI{[\L h]_{abcd}\equiv\delta C_{abcd}\ ,}
where $\delta C_{abcd}$ is the first-order variation of the
Weyl tensor under the trace-free metric variation $\delta
g_{ab}=h_{ab}$:
\eqnn\eVarC
$$\eqalignno{\delta C_{abcd} &= \{ \half\nabla_d\nabla_a h_{bc}
+ \hbox{$1\over4$}g_{da}(-\Box h_{bc} + \nabla_b\nabla^eh_{ce}
+ \nabla_c\nabla^eh_{be})
- \hbox{$1\over12$}g_{da}g_{bc}\nabla^e\nabla^fh_{ef} \cr &\qquad
+ \hbox{$1\over4$}h_{da}\Phi_{bc}
+ \hbox{$1\over4$}g_{da} (h_b{}^e\Phi_{ec} + 2h_c{}^e\Phi_{eb})
- \hbox{$1\over6$}g_{da}g_{bc}h^{ef}\Phi_{ef} \cr &\qquad
+ \hbox{$1\over12$}Rg_{da}h_{bc} + \hbox{$1\over4$}C_{cdae}h^e{}_b
- \half g_{da}h^{ef}C_{cebf}\}  \cr &\quad
- \{a\leftrightarrow b\} - \{c\leftrightarrow d\}
+ \{a\leftrightarrow b, c\leftrightarrow d\}\ .&\eVarC}$$
If the domain of $\L$ is enlarged to include trace perturbations,
we have the relation $(\L\Omega^2g)_{abcd}=\Omega^2C_{abcd}$ for
any non-zero suitably differentiable function $\Omega$. This
follows from the conformal properties of the Weyl tensor. A
special case of this is the following expression for the Weyl
tensor:
\eqn\eLg{C_{abcd}=[\L g]_{abcd}\ .}
As can be seen, the operator $\L$, and consequently $\D_1$,
is a rather complicated second-order differential operator. It
is more compactly expressed in terms of spinors, as we will
show below for the case of $\D_1$ acting on trace-free metric
variations. The kernel of $\D_1$ is the set of metric
perturbations which preserve the self-duality of the Weyl tensor.

Lastly, $\D_2$ is identically the zero operator.

In the spinor notation introduced earlier, $\xi_a$ may be
rewritten as $\xi_{A A^\prime}$. It is illuminating to check
that the degrees of freedom correspond. $\xi_a$ has four
independent components and the dimension of $\Omega_{11}$ is
$(1+1)(1+1)=4$, as from \eDimOmega. The trace-free symmetric
metric variation $h_{ab}$ may be written as $h_{ABA^\prime
B^\prime}\in\Omega_{22}$, and indeed $\dim\Omega_{22}=9$ as
expected.

In terms of spinors, the orthogonal decomposition of the Weyl
tensor \eWeylTensor\ into its self-dual and anti-self-dual
parts reads \rPR
\eqn\COrthogDecomp{C_{abcd}=W_{A^\prime B^\prime C^\prime
D^\prime}\epsilon_{AB}\epsilon_{CD} + W_{ABCD}\epsilon_{A^\prime
B^\prime}\epsilon_{C^\prime D^\prime}\ .}
$W_{A^\prime B^\prime C^\prime D^\prime}\in\Omega_{04}$ is the
self-dual part, while $W_{ABCD}\in\Omega_{40}$ is the
anti-self-dual part. Each of these terms is totally symmetric
in its indices, and has 5 independent components. Together,
they make up the 10 independent components of $C_{abcd}$.

Thus in terms of spinor bundles, the instanton deformation
complex \eSingerCplxI\ may be written as
\eqn\eSingerCplxII{0 \ {\buildrel\D_{-1}\over\longrightarrow}
\ \Omega_{11}
\ {\buildrel\D_0\over\longrightarrow} \ \Omega_{22}
\ {\buildrel\D_1\over\longrightarrow} \ \Omega_{40}
\ {\buildrel\D_2\over\longrightarrow} \ 0\ .}
Note that the dimension of $\Omega_{11}$ plus that of
$\Omega_{40}$ equals the dimension of $\Omega_{22}$.

The two non-trivial operators in \eSingerCplxII\ may be recast as
\eqn\eSpinorDI{[\D_0\xi]_{ABA^\prime B^\prime}
= \half(\nabla_{AA^\prime}\xi_{BB^\prime}
+ \nabla_{BB^\prime}\xi_{AA^\prime}
+ \nabla_{AB^\prime}\xi_{BA^\prime}
+ \nabla_{BA^\prime}\xi_{AB^\prime})\ ,}
and
\eqn\eSpinorDII{[\D_1h]_{ABCD} = \half
[- \nabla_{(A}{}^{(A^\prime}\nabla_B{}^{B^\prime)}
+ \Phi_{(AB}{}^{A^\prime B^\prime}\ ]\
h_{CD)A^\prime B^\prime}\ ,}
where the round brackets in the latter expression indicate
symmetrisation of the indices between them. It is equal to
$\delta W_{ABCD}$ \rTG, the change in the anti-self-dual Weyl
tensor under the trace-free metric variation $\delta g_{ab}
=h_{ab}\equiv h_{ABA^\prime B^\prime}$.

We can also introduce the adjoint operators $\D^\ast_0:
\Omega_{22}\rightarrow\Omega_{11}$ and $\D^\ast_1:\Omega_{40}
\rightarrow\Omega_{22}$ via the relations
\eqn\eAjointI{\int_M \hbox{d}^4x\ \sqrt g\ h_{ABA^\prime B^\prime}
[\D_0\xi]^{ABA^\prime B^\prime} = \int_M \hbox{d}^4x\ \sqrt g\
[\D^\ast_0h]_{AA^\prime}\xi^{AA^\prime}\ ,}
and
\eqn\eAjointII{\int_M \hbox{d}^4x\ \sqrt g\ \chi_{ABCD}
[\D_1h]^{ABCD} = \int_M \hbox{d}^4x\ \sqrt g\
[\D^\ast_1\chi]_{ABA^\prime B^\prime}h^{ABA^\prime B^\prime}\ .}
Explicitly, they are given by
\eqn\eAdjointIII{[\D^\ast_0h]_{AA^\prime}=-2\nabla^{B
B^\prime}h_{ABA^\prime B^\prime}\ ,}
and
\eqn\eAdjointIV{[\D^\ast_1\chi]_{ABA^\prime B^\prime}
=\half[-\nabla^C{}_{(A^\prime}\nabla^D{}_{B^\prime)}
+ \Phi^{CD}{}_{A^\prime B^\prime}\ ]\ \chi_{ABCD}\ .}

\subsec{Ellipticity of the deformation complex}

Before proceeding, we have to check that the instanton
deformation complex \eSingerCplxII\ is an {\it elliptic
complex}. It is a complex because at each step of the sequence,
the image of $\D_{i-1}$ is contained in the kernel of $\D_i$,
i.e., $\D_i\D_{i-1}=0$. In particular, $\D_1\D_0=0$ because the
self-duality of Weyl tensor is preserved under a trace-free
metric perturbation of the form \eLongitudinalP, corresponding
to an unphysical coordinate transformation. Indeed it may be
verified explicitly from the two expressions \eSpinorDI\ and
\eSpinorDII\ via use of the Bianchi identity that
\eqn\eDoneDzero{[\D_1\D_0\xi]_{ABCD}=-\nabla_{EE^\prime}
W_{ABCD}\xi^{EE^\prime}+W_{(ACD}{}^E\nabla_{B)B^\prime}\xi_E
{}^{B^\prime}\ .}
This vanishes {\it on-shell}, i.e., when the Weyl tensor is
self-dual: $W_{ABCD}=0$.

To prove ellipticity, we must first introduce the symbols of the
operators $\D_0$ and $\D_1$. The symbol $\tilde\D$ of an operator
$\D$ is obtained by replacing all partial derivatives in $\D$
with covectors $k_a$, and retaining only the highest order terms
in $k_a$ \rEGH. Thus the symbols of $\D_0$ and $\D_1$ are, in
terms of tensors,
\eqn\eSymI{[\tilde\D_0(k)\xi]_{ab}=k_a\xi_b+k_b\xi_a
-\half g_{ab}k_c\xi^c\ ,}
and
\eqnn\eSymII
$$\eqalignno{[\tilde\D_1(k)h]_{abcd} &= \{ \half k_dk_a h_{bc}
+ \hbox{$1\over4$}g_{da}(-k^2h_{bc}
+ k_bk^eh_{ce} + k_ck^eh_{be}) \cr &\qquad
- \hbox{$1\over12$}g_{da}g_{bc}k^ek^fh_{ef}\} \cr &\quad
- \{a\leftrightarrow b\} - \{c\leftrightarrow d\}
+ \{a\leftrightarrow b, c\leftrightarrow d\}\ .&\eSymII}$$

A way to think of the resulting symbol sequence:
\eqn\eSymSeq{0 \ \longrightarrow\ \Omega_{11}
\ {\buildrel\tilde\D_0\over\longrightarrow} \ \Omega_{22}
\ {\buildrel\tilde\D_1\over\longrightarrow} \ \Omega_{40}
\ \longrightarrow \ 0\ ,}
is that it represents the special case of the full deformation
complex \eSingerCplxII\ when all curvature vanishes, i.e., in
flat space. In such a situation, it is easy to check that
\eqn\eSymIISymI{\tilde\D_1(k)\tilde\D_0(k)=0\ ,}
for $k\neq0$.

The instanton deformation complex \eSingerCplxII\ is elliptic
if its symbol sequence is exact \rEGH:
\eqn\eSymExact{\im\tilde\D_0(k)=\ker\tilde\D_1(k)\ ,}
for $k\neq0$. Since \eSymIISymI\ already implies $\im\tilde\D_0(k)
\subseteq\ker\tilde\D_1(k)$, ellipticity requires
\eqn\eSymExacti{\ker\tilde\D_1(k)\subseteq\im\tilde\D_0(k)\ .}
This condition states that, if the vanishing Weyl tensor of {\it
flat space} is left invariant under a trace-free metric
perturbation $h_{ab}$:
\eqn\eClosed{[\D_1h]_{abcd} = 0\ ,}
then it must follow that $h_{ab}$ is a trivial coordinate
transformation:
\eqn\eExact{h_{ab}=[\D_0\xi]_{ab}\ ,}
for some vector field $\xi^a$. (This is the analogue of the
Poincar\'e lemma for the de~Rham complex---where a closed
form in flat space implies that it is exact.)

Now it can be shown that the map $\tilde\D_0(k)$ is injective,
while $\tilde\D_1(k)$ is surjective, for all $k\neq0$ \rKK. Since
$\Omega_{11}$ is four-dimensional, it follows that the image
of $\tilde\D_0(k)$ forms a four-dimensional subspace of the
nine-dimensional vector space $\Omega_{22}$. On the other
hand, since $\Omega_{40}$ is five-dimensional, the domain of
$\tilde\D_1(k)$ in $\Omega_{22}$ must be at least five-dimensional.
Because of \eSymIISymI, the non-trivial domain of $\tilde\D_1(k)$
is exactly five-dimensional. We have the situation as depicted
in Fig.~1.

%
%
\topinsert
$$\vbox{\offinterlineskip
\def\vline{\vrule width1pt}
\halign{&#\cr
&\multispan1\leaders\hrule height1pt\hfill&&\multispan1\hrulefill&
  &\multispan1\leaders\hrule height1pt\hfill&\cr

&\vline height1cm\hfil\vline&&&&\vline\hfil\vline&\cr

&\vline\quad4\quad\vline&&\hskip1cm$\tilde\D_0(k) \longrightarrow$
  \hskip1cm&&\vline\quad4\quad\hfil\vline&\cr

&\vline height1cm\hfil\vline&&&&\vline\hfil\vline&\cr

&\multispan1\leaders\hrule height1pt\hfill&&\multispan1\hrulefill&
  &\multispan1\leaders\hrule height1pt\hfill&
  &\multispan1\hrulefill&
  &\multispan1\leaders\hrule height1pt\hfill&\cr

&&&&&\vline height1.3cm\hfil\vline&&&
  &\vline height1.3cm\hfil\vline&\cr

&&&&&\vline\quad5\quad\hfil\vline&
  &\hskip1cm$\tilde\D_1(k) \longrightarrow$\hskip1cm&
  &\vline\quad5\quad\hfil\vline&\cr

&&&&&\vline height1.3cm\hfil\vline&&&
  &\vline height1.3cm\hfil\vline&\cr

&&&&&\multispan1\leaders\hrule height1pt\hfill&
  &\multispan1\hrulefill&
  &\multispan1\leaders\hrule height1pt\hfill&\cr
\noalign{\bigskip}
&$\hfil\Omega_{11}\hfil$&&&&$\hfil\Omega_{22}$\hfil&&&
  &$\hfil\Omega_{40}$&\cr
\noalign{\bigskip\medskip}
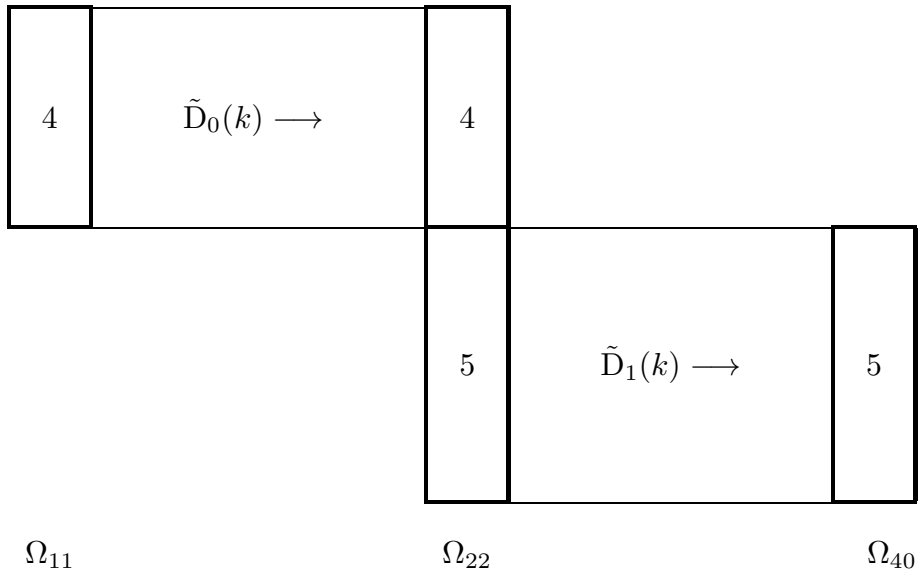
&\multispan{10} \hfil Fig.~1.\quad Dimensions and mappings
  of the symbol sequence \hfil\cr
\noalign{\bigskip}
}}$$
\endinsert
%
%

The proof of \eSymExacti\ then follows immediately. Suppose
$h_{ab}$ is a non-trivial metric perturbation which leaves the
Weyl tensor invariant. Since it does not belong to the image of
$\tilde\D_0(k)$, it must reside in the non-trivial five-dimensional
domain of $\tilde\D_1(k)$. But this leads to a contradiction
as $h_{ab}$ is supposed to leave the Weyl tensor invariant.
Hence the instanton deformation complex is indeed elliptic.

We have seen that $\Omega_{22}$ of the symbol sequence \eSymSeq\
may be decomposed into two orthogonal subspaces, one of which
whose elements may be written as $\tilde\D_0(k)$ of some
$(1,1)$-field, and the other whose elements may be expressed
as $\tilde\D^\ast_1(k)$ acting on some $(4,0)$-field. Such a
decomposition also holds for the full deformation complex, and
this is the generalisation of the Hodge-decomposition theorem
for the de~Rham complex.

Let us introduce the associated Laplacians of the deformation
complex, defined by $\triangle_{(i)}\equiv\D_{i-1}\D^\ast_{i-1}+
\D_i^\ast\D_i$, i.e.,
\eqn\eLaplacian{
\eqalign{\triangle_{(0)} &= \D_0^\ast \D_0\ ,\cr
\triangle_{(1)} &= \D_0\D_0^\ast + \D_1^\ast \D_1\ ,\cr
\triangle_{(2)} &= \D_1\D_1^\ast\ .}}
$\triangle_{(0)}$ is a second-order differential operator
defined on $(1,1)$-fields; $\triangle_{(1)}$ is a combination
of a second-order differential operator and a fourth-order one,
which acts on $(2,2)$-fields; while $\triangle_{(2)}$ is a
second-order differential operator acting on $(4,0)$-fields.
Of course, physically, there is here a mismatch of dimensions.
This may be resolved by implicitly introducing some arbitrary
scale factor with dimension inverse length squared, into the
$\D_0$ terms of \eLaplacian. As a result, $\D_0$ will have
dimensions inverse length squared, and all our Laplacians
will be ``fourth-order'' differential operators. They are
elliptic operators because the deformation complex is elliptic.
Note also that these Laplacians look nothing like the usual ones
arising from the de~Rham complex \rCD. It may be possible to
formulate a relationship between them, however.

The decomposition theorem for the deformation complex may now
be stated. Denote $\Omega_{(0)}\equiv\Omega_{11}$, $\Omega_{(1)}
\equiv\Omega_{22}$ and $\Omega_{(2)}\equiv\Omega_{40}$. If
$\omega_i\in\Omega_{(i)}$, then it can be uniquely decomposed into
the sum
\eqn\eHodge{\omega_i=\D_{i-1}\alpha_{i-1}+\D^\ast_i\beta_{i+1}
+\gamma_i\ ,}
where $\alpha_{i-1}$ is some element of $\Omega_{(i-1)}$, $\beta
_{i+1}$ is some element of $\Omega_{(i+1)}$, and $\gamma_i$ is
an element of $\Omega_{(i)}$ satisfying $\triangle_{(i)}\gamma
_i=0$. The first term is called the exact part of $\omega_i$,
the second term its coexact part, and the last term the harmonic
part. Note that the harmonic part of $\omega_i$ satisfies
\eqn\eHarmonic{\D_i\gamma_i=0\ ,\qquad \D^\ast_{i-1}\gamma_i=0\ .}

Before leaving this subsection, let us make one more remark.
As we have seen, the dimension of $\Omega_{22}$ equals the sum of
the dimensions of $\Omega_{11}$ and $\Omega_{40}$. This suggests
that we can rewrite the deformation complex \eSingerCplxII\ as
\eqn\eSingerCplxIII{\Omega_{22}\ {\buildrel\T\over\longrightarrow}
\ \Omega_{11}\oplus\Omega_{40}\ ,}
where the operator $\T$ is defined by $\T=\D_0^\ast\oplus\D_1$,
and its adjoint is $\T^\ast=\D_0\oplus\D^\ast_1$. This form of
the deformation complex maps trace-free metric variations at a
point to those which are trivial, and to those which deform the
Weyl tensor.

\subsec{The index of the deformation complex}

As we have showed, the operators $\D_i$ of the deformation
complex satisfy $\D_i\D_{i-1}=0$. This implies that $\im\D_{i-1}
\subseteq\ker\D_i$, but unlike its symbol sequence, equality
between the sets does not hold in general. We may then define
the {\it cohomology groups} of the deformation complex to be \rEGH
\eqn\eCohomologyGps{H^i\equiv{\ker\D_i\over\im\D_{i-1}}\ ,}
which consists of elements of the kernel of $\D_i$, identified
if they differ by an element in the image of $\D_{i-1}$. They
may be alternatively written as
\eqn\eCohomologyGpsI{\eqalign{H^i&={\ker\D_i\cap
\ker\D^\ast_{i-1}} \cr&= \ker\triangle_{(i)}\ ,}}
and so there is an isomorphism between $H^i$ and the harmonic
subspace of $\Omega_{(i)}$. These cohomology groups are finite
dimensional, and we set $h^i\equiv\dim H^i$. The {\it index}
of the deformation complex is defined to be
\eqn\eIndex{\hbox{index}\equiv h^0-h^1+h^2\ ,}
and it is a topological quantity which we will calculate in the
next subsection.

Let us now examine these three cohomology groups.

The image of $\D_{-1}$ consists of a single point, so it is
zero-dimensional. Hence, $H^0=\ker\D_0$ is the conformal
group of $M$. On a four-dimensional compact manifold $M$,
the dimension of the conformal group can be at most 15, which
corresponds to a conformally flat manifold such as the
four-sphere $S^4$. This includes both the number of isometries,
which are generated by vector fields $\xi^a$ satisfying
\eqn\eIsometry{\nabla_a\xi_b + \nabla_b\xi_a = 0\ ;}
and the number of non-isometric or proper conformal
transformations satisfying \eConformKV, for $\nabla_a\xi^a\neq0$.
The dimension of the isometry group is at most 10, leaving a
maximum of 5 independent proper conformal transformations on $M$.

A manifold $M$ is called Einstein if it satisfies
\eqn\eEinstein{R_{ab}=\Lambda g_{ab}\ ,}
for some real value of the cosmological constant $\Lambda$.
A theorem of Yano \ref\rYano{K.~Yano, Integral Formaulas in
Riemannian Geometry (Marcel Dekker, New York, 1970)}\ states
that if $M$ is a compact Einstein manifold and if it admits an
infinitesimal proper conformal transformation, then $M$ is
isometric to the four-sphere $S^4$. Hence if $M$ is Einstein
but not isometric to $S^4$, it admits no proper conformal
transformations, and $h^0\leq10$.

Another theorem of Yano \rYano\ states that there are no
conformal Killing vectors on a manifold with negative Ricci
scalar, so that $h^0$ vanishes.

$H^1$ consists of self-dual variations of the Weyl tensor
$(\ker\D_1)$ factored out by trace-free metric fluctuations
of the form of $\im\D_0$, as given in \eLongitudinalP. The
dimension of this space $h^1$ is the number of independent
non-trivial self-dual deformations that can be made around the
self-dual Weyl tensor of $M$. This is precisely the quantity
$\varpi$ given by \eDimMSp, which we are going to find an
expression for.

By \eCohomologyGpsI, we may alternatively write
\eqn\eCohomologyI{H^1=\ker\D^\ast_0\cap\ker\D_1\ .}
The physical interpretation of this is that instead of modding
out trivial metric variations $h_{ab}=[\D_0\xi]_{ab}$, we are
fixing the gauge to be $[\D^\ast_0h]_a=0$. Thus $h^1$ counts the
number of independent solutions to
\eqn\eCohomologyII{\D^\ast_0\psi=0\ ,\qquad \D_1\psi=0\ ,}
for $\psi\in\Omega_{22}$.

$\ker\D_2$ is the whole of the space $\Omega_{40}$. Hence $H^2$
is the subspace of $\Omega_{40}$ orthogonal to the mapping $\D_1$,
or equivalently just $\L$. This cohomology group consists of
fields belonging to the harmonic subspace of $\Omega_{40}$,
i.e., the kernel of $\D^\ast_1$. The dimension of $H^2$ can
be at most 5, corresponding to that of $\Omega_{40}$ itself.

This number in fact vanishes if $R>0$, as we will now show for
the {\it Einstein} case. Any element of $H^2$ belongs to the
kernel of $\triangle_{(2)}$. In the Einstein case, $\triangle
_{(2)}$ may be explicitly evaluated from \eSpinorDII\ and
\eAdjointIV\ to read
\eqn\eTriangleII{[\D_1\D^\ast_1\chi]_{ABCD}=\hbox{$1\over16$}
(-\Box+\hbox{$8\over3$}\Lambda)(-\Box+2\Lambda)\chi_{ABCD}\ .}
This operator is positive definite if $R=4\Lambda>0$. In such
a case, the kernel of $\triangle_{(2)}$ has to be trivial, and
so $h^2$ vanishes. We will show below via an example that, in
the case of a manifold with a self-dual Riemann tensor, $h^2$
counts the number of covariantly constant objects in $\Omega_{40}$.

\subsec{Calculation of the index}

The Atiyah--Singer index theorem may now be applied to the
instanton deformation complex. It is a formula which simply
relates the index of an elliptic complex to the topological
properties (twisting) of the bundles $\Omega_{mn}$ associated
with it. On a four-dimensional Riemannian manifold $M$, it
reads \rShan
\eqn\eIndexThm{\hbox{index}=\int_M {{\ch(\bigoplus (-1)^iE_i) \
\td(TM\otimes \C)}\over{\e(TM)}}\ .}
The integral over $M$ indicates that we extract the 4-forms in
the integrand, and integrate them over $M$. $TM$ is the tangent
bundle of $M$, and $TM\otimes\C$ its complexification. The $E_i$'s
are the vector bundles over $M$ associated with the elliptic
complex. In our case, the Whitney sum over the bundles as
indicated in the index theorem is
\eqn\eWhitneySum{\bigoplus (-1)^iE_i=\Omega_{11}\ominus
\Omega_{22}\oplus\Omega_{40}\ .}
ch, td and e are the Chern character, Todd class and Euler class
of the various vector bundles involved.

A very useful tool used in evaluating the index theorem is the
splitting principle, which allows us to treat any vector bundle
as a direct sum of one-dimensional line bundles. For example,
it implies that
\eqn\eChernChar{\eqalign{\ch(\bigoplus (-1)^iE_i) &=
\ch(\Omega_{11}\ominus \Omega_{22}\oplus\Omega_{40}) \cr
&= \ch(\Omega_{11})-\ch(\Omega_{22})+\ch(\Omega_{40})\ .}}
One can also use it to derive the expressions \rShan
\eqn\eEulerClass{\e(TM)=\prod_{i=1}^2x_i=x_1x_2\ ,}
\eqn\eToddClass{\td(TM\otimes\C) = \prod_{i=1}^2{-x_i^2 \over
(1-\e^{-x_i})(1-\e^{x_i})}\ ,}
where $x_i=\lambda_i/2\pi$ are two-forms, for $\lambda_1$ and
$\lambda_2$ the two independent eigenvalues of the curvature
two-form \eRiemann. The latter expression can be simplified to
give
\eqn\eToddClassI{\td(TM\otimes\C) = 1-\hbox{$1\over 12$}\p_1\ ,}
where $\p_1=x_1^2+x_2^2$ is the first Pontryagin class of $TM$.

R\"omer \ref\rRomer{H.~R\"omer, Phys.~Lett. B83 (1979) 172}\
has calculated the Chern character of $\Omega_{mn}$ to
fourth-order in $x_i$, and he finds that
\eqn\eChernMNi{\eqalign{\ch(\Omega_{mn}) &=A_mA_n
+ \half A_mB_n(x_1+x_2)^2
+ \half A_nB_m(x_1-x_2)^2 \cr &\quad
+ \hbox{$1\over4$}B_mB_n(x_1^2-x_2^2)^2
+ \hbox{$1\over24$}A_mC_n(x_1+x_2)^4
+ \hbox{$1\over24$}A_nC_m(x_1-x_2)^4\ ,}}
where
$$A_n=n+1\ ; \quad B_n=\sum_{k=0}^{n}(k-\half n)^2\ ;
\quad C_n=\sum_{k=0}^{n}(k-\half n)^4.$$
One can then readily show that
\eqn\eChernMNii{\ch(\Omega_{11})-\ch(\Omega_{22})
+\ch(\Omega_{40}) = -10x_1x_2
+ \hbox{$15\over2$}x_1^2x_2^2
- \hbox{$17\over3$}x_1x_2(x_1^2+x_2^2)\ .}
Substituting all this into the index theorem gives
\eqnn\eIndexI
$$\eqalignno{\hbox{index} &=\int_M(-10+\hbox{$15\over2$}x_1x_2-
   \hbox{$17\over3$}\p_1)(1-\hbox{$1\over12$}\p_1) \cr
&= \int_M (\hbox{$15\over2$}\e-\hbox{$29\over6$}\p_1) \cr
&= \half(15\chi-29\tau)\ .
&\eIndexI}$$
Similarly, considering the case of opposite orientation
($\Omega_{04}$ instead of $\Omega_{40}$) gives the index to be
\eqn\eIndexII{{\rm index}=\half(15\chi+29\tau)\ .}
Bear in mind that a conformally self-dual manifold would have
Weyl tensor belonging to $\Omega_{04}$, and have a positive
signature $\tau$. On the other hand, a conformally anti-self-dual
manifold would have Weyl tensor belonging to $\Omega_{40}$, and
have negative $\tau$. Hence the index is actually given by
\eqn\eIndexIII{{\rm index}=\half(15\chi-29|\tau|)\ .}

\nref\rAPS{M.F.~Atiyah, V.K.~Patodi and I.M.~Singer,
Math.~Proc.~Camb.~Phil.~Soc. 77 (1975) 43; 78 (1975) 405;
79 (1976) 71}
\nref\rRomerI{A.J.~Hanson and H.~R\"omer, Phys.~Lett. B80 (1978)
58\semi G.W.~Gibbons, C.N.~Pope and H.~R\"omer, Nucl.~Phys.
B157 (1979) 377}

Finally, recall that most of the known gravitational instantons
are, in fact, non-compact. However, it is possible to treat them
as compact manifolds with boundaries which recede to infinity.
The presence of a boundary will modify the value of the index
derived above. The new index may be calculated via an application
of the Atiyah--Patodi--Singer index theorem \rAPS, which is the
extension of the Atiyah--Singer index theorem to manifolds with
boundary, to the deformation complex. In fact, R\"omer and
collaborators \refs{\rRomer, \rRomerI}\ have calculated such
boundary corrections of certain index theorems for several
gravitational instantons with boundary. In particular, they have
treated the Eguchi--Hanson and other asymptotically locally
Euclidean instantons, whose boundaries are $S^3$ identified
under discrete subgroups of SU(2). One could try to generalise
their calculations to the instanton deformation complex.

\subsec{Moduli space of gravitational instanton metrics}

To summarise, we have so far obtained the relation defined on a
self-dual manifold $M$:
\eqn\eHI{\varpi=\half(29|\tau|-15\chi) + \hbox{dim(conformal
group of $M$)} + h^2,}
where $h^2$ is some ``small'' correction which vanishes if
$M$ is Einstein and has positive Ricci scalar.

Now consider the space of all metrics, modded out by conformal
and coordinate transformations. Given a manifold $M$, we define
the {\it gravitational moduli space} of $M$ to be that subset
of the former space consisting of metrics on $M$ which are
conformally self-dual.

Given a particular conformally self-dual metric $g_0$ on $M$,
\eHI\ counts the number of independent tangent vectors to $g_0$
in the moduli space. This is because in our derivation of \eHI,
we have assumed that the metric variations are infinitesimal.

Because the last two terms of \eHI\ depend on the metric,
``integrating'' $g_0$ to give a local manifold structure for
the moduli space is in general not well defined. This means that
the moduli space may have singularities around $g_0$. On the
other hand, if a local manifold structure existed around $g_0$,
its dimension would be given by $\varpi$.

Even if there are no obstructions to defining a local manifold
structure around $g_0$, the moduli space still may not exist
globally as a manifold. This could happen if there existed
another conformally self-dual metric $g_1$ on $M$, which is
``very different'' from $g_0$. The number of independent tangent
vectors at $g_1$ could be different from that at $g_0$. Thus
the moduli space might consist of disconnected compact manifolds
with different dimensions.

The gravitational moduli space is much more well behaved if the
two non-topological terms of \eHI\ are not present. Let us give a
mathematical characterisation of this, which follows the analysis
applied to the Yang--Mills moduli space \rAHS. Consider the
deformation complex in the form \eSingerCplxIII. For a given
metric $g$ on $M$, tangent vectors to $g$ belong to the
cohomology group $H^2$. This is equivalently the kernel
of T. If T is a surjective map, then the implicit function
theorem can be invoked to deduce that the kernel of T is smooth
near $g$. T is surjective if the kernel of $\T^\ast$ vanishes:
\eqn\eKerTStar{\ker\T^\ast=\ker\D_0\oplus\ker\D^\ast_1
=H^0\oplus H^2=0\ .}
Thus the moduli space exists as a smooth manifold around $g$ if
$h^0$ and $h^2$ both vanish.

An analogous analysis of the Yang--Mills case \rAHS\ may then be
applied here to show that these local manifolds can be patched
together to give a global manifold structure to the moduli
space. This gravitational moduli space will have a topology
induced by the space of conformally and physically inequivalent
metrics, and whose dimension is equal to $\half(29|\tau|-15\chi)$.
It will in general be a smooth manifold, except perhaps for
isolated instances of singularities.

$\varpi$ is called the virtual dimension of the moduli space.
It is virtual because it does not imply the existence of
conformally self-dual metrics when $\varpi>0$. What it means is
that if such a metric exists, then it would have a moduli space
of dimension $\varpi$ around it, or more generally $\varpi$
tangent vectors at that point in the moduli space.

\subsec{Some examples}

Presently, we do not know of any examples of compact conformally
self-dual manifolds which are not conformally Einstein. Thus the
following theorem of Hitchin \ref\rBesse{A.L.~Besse, Einstein
Manifolds (Springer--Verlag, Berlin, 1987)}\ is of some importance
to us:

Let $M$ be a compact conformally self-dual Einstein manifold. Then

(1)\enskip If $R>0$, $M$ is either isometric to the four-sphere
$S^4$, or to the complex projective two-space $\C P^2$, with their
standard metrics;

(2)\enskip If $R=0$, $M$ is either flat (for example, the flat
metric on the four-torus $T^4$) or its universal covering
is the K3 surface with the Calabi--Yau metric.

Bearing this uniqueness result in mind, let us now compute
$\varpi$ for the manifolds listed above.

The standard constant curvature metric on $S^4$ is conformally
flat, so it has a vanishing Weyl tensor which is trivially
self-dual. It has $\chi=2$ and $\tau=0$, giving the index to be
15. The conformal group of this manifold is 15-dimensional,
and $h^2=0$ since $R>0$. Substituting these values into \eHI\
yields the value zero, which shows that the constant curvature
metric on $S^4$ admits no conformally self-dual deformations
except for a scale.

It can be shown \rAHS\ that the standard metric on $S^4$ is the
unique conformally self-dual one. Hence the gravitational moduli
space of $S^4$ consists of a single point.

The standard metric on $\C P^2$ is the Fubini--Study metric \rEGH,
which has a self-dual Weyl tensor. For this case $\chi=3$ and
$\tau=1$, and so the index is 8. Its conformal group is
8-dimensional. Again $R>0$, so $h^2$ vanishes. Thus applying the
formula \eHI\ shows that there are no conformally self-dual
deformations of the standard metric on $\C P^2$, apart from a
scale.

In this case, we cannot rule out the possibility of finding other
examples of conformally self-dual metrics with topology $\C P^2$.

The K3 surface with the Calabi--Yau metric has a self-dual
Riemann tensor, so it is conformally self-dual and $R$ vanishes.
Now $\chi=24$ and $\tau=16$, giving an index of $-52$. Its
conformal group is zero-dimensional. In this case, $h^2$ is
non-vanishing and we will now calculate its value.

The number of anti-self-dual harmonic 2-forms on K3 is \rEGH
\eqn\eBIIMinus{b_2^-=\half(-\tau+\chi-2)=3\ .}
Such a harmonic 2-form which is anti-self-dual satisfies \rCD
\eqn\eHarmonicEqn{\eqalign{\triangle F_{AB}&=
-\Box F_{AB}+\hbox{$4\over3$}\Lambda F_{AB}+2W_{ABCD}F^{CD} \cr
&=0\ .}}
But in the present case, this equation reduces down to simply
$\Box F_{AB}=0$. Now $F_{AB}$ may be decomposed into a symmetric
product of two principle spinors \rPR: $F_{AB}=\epsilon^{(1)}_{(A}
\epsilon^{(2)}_{B)}$. The fact that there are exactly three of
these objects on K3 implies that there are precisely two
covariantly constant unprimed principle spinors $\iota_A$ and
$o_B$ that exist on K3 \ref\rHP{S.W.~Hawking and C.N.~Pope,
Nucl.~Phys. B146 (1978) 381}. The three anti-self-dual harmonic
2-forms are then $\iota_{(A} \iota_{B)}$, $\iota_{(A}o_{B)}$
and $o_{(A}o_{B)}$.

{}From the two principle spinors, we can also construct five
covariantly constant objects in $\Omega_{40}$: $\iota_{(A}
\iota_B\iota_C\iota_{D)}$, $\iota_{(A}\iota_B\iota_Co_{D)}$,
$\iota_{(A}\iota_Bo_Co_{D)}$, $\iota_{(A}o_Bo_Co_{D)}$ and
$o_{(A}o_Bo_Co_{D)}$. These five quantities clearly satisfy
the equation \eTriangleII, and so must belong to $H^2$. Hence,
$h^2=5$ for K3. Applying the formula \eHI\ shows that there are
57 independent conformally self-dual deformations of K3, aside
from a scale. Altogether, the K3 instanton with the Calabi--Yau
metric has 58 parameters.

\newsec{Topological conformal gravity}

Having developed a deformation theory of gravitational instantons
in the previous section, we are now ready to tackle the theory of
topological gravity. Starting with the classical action of
topological conformal gravity, we analyse its symmetries and
BRST-quantise it by introducing the appropriate ghost fields and
gauge-fixing functions. These ghost fields are of the same nature
as the fields that appear in the deformation complex \eSingerCplxII,
and their interaction terms involve the differential operators
$\D_0$ and $\D_1$ of the deformation complex. We then evaluate
the partition function of the quantum theory, and show how this
produces an invariant which looks very much like the first
Donaldson polynomial of topological Yang--Mills theory.

Throughout this section, we will be closely imitating the way
one goes about in dealing with the topological Yang--Mills case,
particularly the work of Baulieu and Singer \rBS. Indeed, we
will follow their notation as far as naming the fields is
concerned, and {\it not} that of Witten \rTG.

\subsec{Classical theory}

Conformal gravity may be regarded as the theory which arises from
the classical action
\eqn\eClassicalCG{\int_M \hbox{d}^4x\ \sqrt g\ C_{abcd} C^{abcd}.}
It is a higher-derivative gravity theory which is not only
invariant under coordinate and Lorentz transformations, but also
conformal transformations. The theory is power-counting
renormalisable, and it was once hoped that it could describe
gravity at Planck lengths, while reducing to ordinary gravity
at larger distances. However, because of the conformal anomaly and
problems with unitarity, it has since become less fashionable.
A review of conformal gravity and its quantisation may be found
in ref.~\ref\rTseytlin{E.S.~Fradkin and A.A.~Tseytlin, Phys.~Rep.
119 (1985) 233}.

The important property of the action \eClassicalCG\ that we will
need below is that it is minimised by metrics whose Weyl
tensors are (anti-) self-dual, as in \eConformSD\ or \eConformASD.
This follows immediately from the inequality \rSHP
\eqn\eCSquared{(C_{abcd} \pm \ast C_{abcd})^2 \ge 0\ .}
Such manifolds are known as conformally (anti-) self-dual
gravitational instantons, and known examples \rEGH\ include
the standard metrics on flat Euclidean four-space, four-torus,
four-sphere, the Eguchi--Hanson metric, the Taub--NUT metric,
the Fubini--Study metric on $\C P^2$, the Gibbons--Hawking
multi-centre metrics, and the Calabi--Yau metric on K3.

It is clear that the action \eClassicalCG\ is the closest
gravitational analogue to the Yang--Mills functional \eClassicalYM\
available. Indeed, bearing in mind the action for topological
Yang--Mills theory \eClassicalTYM, the classical action for
topological conformal gravity is very similarly,
\eqn\eClassicalTG{\int_M \hbox{d}^4x\ \sqrt g\ C_{abcd}
\ast C^{abcd},}
which up to a factor is the Hirzebruch signature \eTau.
The easiest way to see that \eClassicalTG\ is invariant under
arbitrary variations of the metric is to recognise that it is
equivalent to
\eqn\eClassicalTGi{\int_M \hbox{d}^4x\ \sqrt g\ R_{abcd}
\ast R^{abcd},}
which in turn can be written in terms of forms as
\eqn\eClassicalTGii{\int_M \tr\{R\wedge R\}\ ,}
where the curvature 2-form is given by $R^a{}_b={\rm d}\omega^a{}_b
+\omega^a{}_c\wedge\omega^c{}_b$, for connection 1-forms
$\omega^a{}_b$. Varying \eClassicalTGii\ with respect to
$\omega^a{}_b$ quickly shows that it is invariant under
arbitrary changes in $\omega^a{}_b$, and hence under arbitrary
variations of the metric.

To be more precise about this metric variation, recall from
ref.~\ref\rGP{G.W.~Gibbons and M.J.~Perry, Nucl.~Phys. B146
(1978) 90}\ that any metric perturbation $\delta g_{ab}=h_{ab}$
may be decomposed into its trace-free part $\tilde\phi_{ab}$, and
its trace part $\phi g_{ab}$:
\eqn\ePertOrthogDecomp{h_{ab}=\tilde\phi_{ab}+
\hbox{$1\over4$}\phi g_{ab}\ ,\qquad \tilde\phi^a{}_a=0\ .}
$\phi$ corresponds to conformal rescalings of the metric, which
we will as usual ignore. $\tilde\phi_{ab}$ may be further
decomposed orthogonally into its transverse trace-free part
$\phi_{ab}^{\rm TT}$, and its longitudinal part $\phi_{ab}
^{\rm L}$:
\eqn\eTTandL{\tilde\phi_{ab}=\phi_{ab}^{\rm TT}+\phi_{ab}^{\rm L}
\ ,}
where $\phi_{ab}^{\rm L}$ may be expressed in terms of a vector
field $\xi_a$ as
\eqn\eLVar{\phi_{ab}^{\rm L}=\nabla_a\xi_b + \nabla_b\xi_a -
\half g_{ab}\nabla_c\xi^c,}
and the transverse trace-free
part satisfies
\eqn\eTTVar{\nabla^a\phi_{ab}^{\rm TT}=0\ .}

Hence to characterise an arbitrary metric variation of
\eClassicalTG, consider the quantum metric
fluctuation
\eqn\eMetricPert{g_{ab}=g^{\rm cl}_{ab}+h_{ab}\ ,}
about a fixed classical background $g^{\rm cl}_{ab}$. We project
out the trace part, so that $h_{ab}$ takes the form \eTTandL.
In spinor notation, $h_{ab}=h_{ABA^\prime B^\prime}$ belongs to
$\Omega_{22}$. The longitudinal part of $h_{ab}$ is of the form
$\D_0$ acting on some vector field. Thus, it characterises that
part of $\Omega_{22}$ which is exact, by our decomposition
theorem of the instanton deformation complex. Furthermore, the
transverse trace-free part of $h_{ab}$ consists of the coexact
sector of $\Omega_{22}$; and \eTTVar\ follows from the property
that $\D_1\D_0=0$:
\eqn\eTTProp{\nabla^bh_{ab}^{\rm TT}
\propto(\D_0^\ast h^{\rm TT})_a=(\D_0^\ast\D_1^\ast\chi)_a=0\ .}
Note that the transverse trace-free also consists of elements
of $\Omega_{22}$ which are harmonic.

Thus, we can write an arbitrary quantum metric fluctuation of
\eClassicalTG\ in the form
\eqn\eTetradVar{h_{ABA^\prime B^\prime}=\rho_{ABA^\prime
B^\prime} + [\D_0\omega]_{ABA^\prime B^\prime}\ .}
$\rho_{ABA^\prime B^\prime}$ belongs to $\Omega_{22}$, while
$\omega_{AA^\prime}$ is an element of $\Omega_{11}$.

\subsec{BRST symmetry and gauge-fixing}

Since the classical action is invariant under the metric variation
\eTetradVar, we must gauge-fix this topological symmetry in order
for the Feynman path integral of \eClassicalTG\ to make sense.
To do so, we will use the BRST quantisation scheme. In the first
step of this procedure, we introduce ghost fields $\psi_{AB
A^\prime B^\prime}$ and $c_{AA^\prime}$ corresponding to variations
$\rho_{ABA^\prime B^\prime}$ and $\omega_{AA^\prime}$ of
\eTetradVar\ respectively. One may think of $c_{AA^\prime}$ as
a ghost which gauges diffeomorphisms of the form \eLVar, while
$\psi_{ABA^\prime B^\prime}$ as a ghost which gauges the
topological symmetry. Both $\psi_{ABA^\prime B^\prime}$
and $c_{AA^\prime}$ have ghost number +1 and are therefore
anti-commuting.

At this stage, observe that $\rho_{ABA^\prime B^\prime}$ in
\eTetradVar\ is defined modulo a term of the form
\eqn\eSecondGenSymm{\rho_{ABA^\prime B^\prime}\sim\rho_{AB
A^\prime B^\prime} + [\D_0\Lambda]_{ABA^\prime B^\prime}\ .}
One can say that the ghost field $\psi_{ABA^\prime B^\prime}$
itself has its own gauge invariance, which we must also
gauge-fix by introducing a second generation ghost field
$\phi_{A A^\prime}$ corresponding to $\Lambda_{A A^\prime}$.
It has ghost number +2, and is commuting.

The BRST version of the transformation \eTetradVar\ is
\eqn\eSTetradVar{\s\,g_{ABA^\prime B^\prime}=
\s\,h_{ABA^\prime B^\prime}=\psi_{ABA^\prime B^\prime}
+ [\D_0c]_{ABA^\prime B^\prime}\ ,}
where s is the BRST operator. Note that s raises the ghost
number of its operand by one unit, and is anti-commuting. The
BRST version of the symmetry \eSecondGenSymm\ is
\eqn\eSPsiVar{\s\,\psi_{ABA^\prime B^\prime}=
[\D_0\phi]_{ABA^\prime B^\prime}\ .}
{}From \eSTetradVar, one could proceed to calculate s acting on
the inverse metric $g^{ab}$, via the condition that $\s\,\{g^{ab}
g_{bc}\}=0$. Another quantity one could then compute is the
action of s on the Christoffel symbol $\Gamma^a_{bc}$. However,
the contribution from these types of terms to the final
gauge-fixed action are of at least third order in the quantum
fields. We may neglect such higher order contributions since
in this theory, like topological Yang--Mills theory, the
semi-classical limit is exact \rTQFT. Thus in what follows, we
only write down the s transformations to lowest order. The BRST
symmetry will still be preserved if one consistently incorporates
higher order terms as necessary. In particular to lowest order,
$\s$ commutes with the operators $\D_0$ and $\D_1$.

Now, because $W_{ABCD}=(\D_1g)_{ABCD}$ if we extend the domain of
$\D_1$ as in \eLg, we have
\eqn\eSWVar{\eqalign{\s\,W_{ABCD} &= [\D_1\s\,g]_{ABCD} \cr
&= [\D_1(\psi+\D_0c)]_{ABCD}\ .}}
Remember that $\D_1\D_0=0$ only when on-shell, when the
Weyl tensor is self-dual.

By construction, s must be a nilpotent operator, i.e., it
satisfies $\s^2=0$. To enforce this, note that we immediately
have from \eSPsiVar\ that
\eqn\eSPhiVar{\s\,\phi_{AA^\prime}=0\ .}
It also follows from \eSPsiVar\ and \eSTetradVar\ that we must
set
\eqn\eSCVar{\s\,c_{A A^\prime}=-\phi_{A A^\prime}\ .}
Also observe that
\eqn\eSSquaredW{\eqalign{\s^2W_{ABCD} &= [\D_1\s\psi+\D_1\D_0
\s c]_{ABCD}\cr &=[\D_1\D_0\phi-\D_1\D_0\phi]_{ABCD}\cr &=0\ .}}
Hence the operator s defined by the above set of transformations
is indeed nilpotent, even when off-shell.

Having introduced the three ghost fields and the BRST operator s,
it is now time to choose the gauge-fixing terms. Corresponding
to the first symmetry of \eSTetradVar, we may impose the choice
of gauge that
\eqn\eSDCondition{W_{ABCD}=0\ ,}
that is, the Weyl tensor be self-dual.

The second symmetry in \eSTetradVar\ and the second generation
one in \eSPsiVar\ are similar in nature, and so would have the
same type of gauge-fixing function. The normal choice of such a
function would have the form $[\D^\ast_0\lambda]_{AA^\prime}=0$
for some field $\lambda_{ABA^\prime B^\prime}\in \Omega_{22}$.
By considering ghost numbers, we would have the two respective
gauge-fixing functions to be\foot{The former is the usual
harmonic or de~Donder gauge condition.}
\eqna\eGaugeFixFun
$$[\D^\ast_0h]_{AA^\prime}=0\ ,
\eqno\eGaugeFixFun a$$
and
$$[\D^\ast_0\psi]_{AA^\prime}=0\ .
\eqno\eGaugeFixFun b$$

Corresponding to the three ghost fields $\psi_{ABA^\prime
B^\prime}$, $c_{A A^\prime}$ and $\phi_{A A^\prime}$, we have
also to introduce their anti-ghost fields $\bar\chi_{ABCD}$,
$\bar c_{A A^\prime}$ and $\bar\phi_{A A^\prime}$. The BRST
transforms of these anti-ghosts are just three Lagrange multiplier
fields, respectively denoted $B_{ABCD}$, $b_{AA^\prime}$ and
$\bar\eta_{A A^\prime}$, for our gauge-fixing functions. In
other words, we have
\eqn\eSGhosts{\eqalign{\s\,\bar\chi_{ABCD}&=B_{ABCD}\ ,\cr
\s\,\bar c_{A A^\prime}&=b_{A A^\prime}\ ,\cr
\s\,\bar\phi_{A A^\prime}&=\bar\eta_{A A^\prime}\ ,}
\qquad\eqalign{\s\,B_{ABCD}&=0\ ,\cr
\s\,b_{A A^\prime}&=0\ ,\cr
\s\,\bar\eta_{A A^\prime}&=0\ .}}

These therefore are the fields that characterise topological
conformal gravity. It should be clear to the reader by now that
spinor indices just form an unnecessary clutter in our equations.
We will work towards an index-free notation, where the various
fields we have introduced are understood to belong to the
appropriate one of the three representations $\Omega_{11}$,
$\Omega_{22}$ or $\Omega_{40}$. For future reference, we list
these fields and their properties in Table 1.

%
%
\topinsert
\def\singleline{\smallskip\hrule\smallskip}
\def\doubleline{\smallskip\hrule\vskip1pt\hrule\medskip}
$$\vbox{
\bigskip
\line{\hfill Table 1.\quad The fields of topological conformal
      gravity  \hfill}
\medskip
\doubleline
\tabskip=1em plus2em minus .5em
\halign to \hsize{$\hfil#\hfil$&\hfil#\hfil&$\hfil#\hfil$&
      $\hfil#\hfil$&$\hfil#\hfil$&$\hfil#\hfil$\cr
&&&$Ghost$&$Dimension$&$Conformal$ \cr
\noalign{\vskip-5pt}
$ Field $& Meaning &$ Representation $&$Number$&$in [Length]$^n$$
&$ Weight\rlap* $\cr
\noalign{\singleline}
h & metric variation & (2,2) & ~~0 & ~~0 & 2 \cr
C & Weyl tensor & (4,0)\oplus(0,4) & ~~0 & -2 & 2 \cr
W_+ & self-dual part of $C$ & (0,4) & ~~0 & -2 & 0 \cr
W_- & anti-self-dual part & (4,0) & ~~0 & -2 & 0 \cr
\noalign{\singleline}
c & diffeomorphism ghost & (1,1) & +1 & +2 & 2 \cr
\bar c & anti-ghost of $c$ & (1,1) & -1 & -2 & k \cr
b & Lagrange multiplier & (1,1) & ~~0 & -2 & k \cr
\noalign{\singleline}
\psi & topological ghost & (2,2) & +1 & ~~0 & 2 \cr
\bar\chi & anti-ghost of $\psi$ & (4,0) & -1 & -2 & 0 \cr
B & Lagrange multiplier & (4,0) & ~~0 & -2 & 0 \cr
\noalign{\singleline}
\phi &ghost for ghost $\psi$ & (1,1) & +2 & +2 & 2 \cr
\bar\phi &anti-ghost of $\phi$ & (1,1) & -2 & -2 & l \cr
\bar\eta & Lagrange multiplier & (1,1) & -1 & -2 & l \cr}
\doubleline
* (with all indices {\it downstairs})\hfil
\medskip}$$
\endinsert
%
%

\subsec{Conformal weights}

Note that in Table 1, we have listed down the conformal weights
of our fields. This is because as its name implies, conformal
gravity is classically a conformally invariant theory and so we
should be careful to check that no conformal anomaly develops
in the quantum theory.

Our convention here follows that of Penrose and Rindler \rPR,
who consider the metric rescaling given by
\eqn\eMetricRescaling{g_{ab}\mapsto\hat g_{ab}=\Omega^2g_{ab}\ ,}
where $\Omega$ is a non-zero suitably differentiable function.
{}From this and the BRST transformations listed in the previous
subsection, we can therefore derive the conformal weights of
all the fields. Note that the choice of conformal weights for
the anti-ghost fields (and hence the Lagrange multiplier fields)
are essentially free, and should be chosen for later consistency.

At this stage, recall from sec.~5.9 of ref.~\rPR\ that the
equation $\nabla^{BB^\prime}\lambda_{ABA^\prime B^\prime}$
transforms as a conformal density (of weight $-2$) if and
only if $\lambda_{ABA^\prime B^\prime}$ has conformal weight
$-2$. Thus, both of the gauge-fixing conditions in
\eGaugeFixFun{}\ are not (locally) conformally invariant,
since $h_{ABA^\prime B^\prime}$ and $\psi_{ABA^\prime B^\prime}$
do not have conformal weight $+2$. If we were to proceed blindly
ahead, our resulting gauge-fixed action may not be conformally
invariant.

Witten was faced a rather similar problem in ref.~\rTG, which he
tried to resolve by introducing a new field with the appropriate
conformal weight to the theory. However, this naturally made
the theory very arbitrary. Soon after, Labastida and Pernici
\rLP\ pointed out that introducing a new field was unnecessary,
and one could obtain conformal invariance by inserting appropriate
powers of $g=\det(g_{ab})$, which has conformal weight $+8$, into
the right places. For example, our gauge-fixing conditions
\eGaugeFixFun{}\ should be modified to read
\eqn\eGaugeFixFunI{[\D^\ast_0(g^{-1/2}h)]_{AA^\prime}=
[\D^\ast_0(g^{-1/2}\psi)]_{AA^\prime}=0\ .}

We will implicitly adopt this procedure, and insert appropriate
powers of $g$ into the appropriate places. They will not be
written down explicitly for notational simplicity. It will
become clear below that the quantum theory indeed preserves
conformal invariance.

\subsec{Evaluation of the partition function}

The time has arrived for the quantisation of \eClassicalTG,
which may be rewritten in our index-free notation as
\eqn\eClassicalTGiii{\int_M \hbox{d}^4x\ \sqrt g\ (W_+^2
- W_-^2)
\ .}
The partition function for topological conformal gravity is
\eqn\eZforQTG{Z=\int{\cal D}X \exp(-I_{\rm GF})\ ,}
where ${\cal D}X$ represents the path integral over the fields
$h$, $c$, $\bar c$, $b$, $\psi$, $\bar\chi$, $B$, $\phi$,
$\bar\phi$ and $\bar\eta$. The gauge-fixed action consists of
the classical action minus an s-exact part:
\eqn\eGFAction{I_{\rm GF}=\int_M \hbox{d}^4x \sqrt g\
\Bigl[\,(W_+^2-W_-^2) - \s\,\{\cdots\}\,\Bigr]\ .}
Since s is nilpotent, $I_{\rm GF}$ is BRST invariant for any
choice of terms in the curly brackets (with vanishing ghost
number). We judiciously choose
\eqn\eSBrackets{\s\,\{\cdots\} = \s\,\{\bar\chi W_- +
\hbox{$1\over8$}\alpha\bar\chi B + \bar c\D^\ast_0h +
\hbox{$1\over4$}\beta\bar cb + \bar\phi\D^\ast_0\psi\}\ ,}
where $\alpha$ and $\beta$ are real gauge parameters. The
first, third and last terms in the brackets have the form
(anti-ghost)$\times$(gauge-fixing condition). s acting on this
then gives terms of the form (Lagrange multiplier)$\times
$(gauge-fixing condition) in the Lagrangian, plus other terms
describing interactions amongst the ghost fields. The second
and fourth terms in the brackets give rise to terms of the
form $\hbox{(Lagrange multiplier)}^2$. When expanded out using
the properties of s, \eSBrackets\ reads
\eqn\eSBracketsI{\eqalign{\s\,\{\cdots\} &= BW_-
+ \hbox{$1\over8$}\alpha B^2 + \bar\chi\D_1\psi
+ \bar\chi\D_1\D_0c + b\D^\ast_0h \cr&\quad
+ \hbox{$1\over4$}\beta b^2 + \bar c\D^\ast_0\psi
+ \bar c\D^\ast_0\D_0c + \bar\eta\D^\ast_0\psi
+ \bar\phi\D^\ast_0\D_0\phi\ .}}

Firstly, let us choose the gauge $\alpha=\beta=1$. Since $B$ is
not a dynamical field, it may be eliminated via its equation
of motion $B=-4W_-$ to give
\eqn\eBGaussInteg{BW_- + \hbox{$1\over8$} B^2\sim -2W_-^2\ .}
This term adds with the classical action of topological conformal
gravity to give a term of the form $W_+^2+W_-^2=C^2$ in the
Lagrangian. This indeed is just the classical action of {\it
ordinary} conformal gravity.

Similarly, we can eliminate the $b$ field via its equation of
motion $b=-2\D^\ast_0h$ to yield a term of the form
\eqn\ebGaussInteg{b\D^\ast_0h + \hbox{$1\over4$}b^2 \sim
-(\D^\ast_0h)^2 \sim h\D_0\D^\ast_0h\ .}
One can also absorb the term $\bar c\D^\ast_0\psi$ into the term
$\bar\eta\D^\ast_0\psi$ via a redefinition of the field $\bar\eta$.
Hence the total gauge-fixed action becomes
\eqn\eGFActionI{I_{\rm GF}=\int_M \hbox{d}^4x
\sqrt g\ (C^2 - h\D_0\D^\ast_0h - \bar\chi\D_1\psi
- \bar\chi\D_1\D_0c - \bar\eta\D^\ast_0\psi - \bar c\D^\ast_0\D_0c
- \bar\phi\D^\ast_0\D_0\phi)\ .}

We are now ready to evaluate the partition function \eZforQTG.
Note that the gauge-fixed action \eGFActionI\ consists of
classical conformal gravity together with quantum ghost fields
that cancel out all the local degrees of freedom in the classical
term. Hence classical minima of this theory are just conformally
self-dual gravitational instantons as discussed earlier, for these
form the dominant contribution to the partition function
\eZforQTG. To characterise the quantum fluctuations around
these instanton solutions, we introduce the quadratic partition
function
\eqn\eZQuad{Z^{(2)}=\int{\cal D}X\ \exp(-I^{(2)})\ ,}
corresponding to the part of the action quadratic in the quantum
fields. The total partition function is then the sum of the
quadratic partition functions $Z^{(2)}$ over gravitational
instantons, whose self-duality enforce the condition that
$\D_1\D_0=0$.

By performing the remaining Gaussian integrals over the
commuting and anti-commuting quantum fields, we can represent
$Z^{(2)}$ as a ratio of determinants of the Laplacians involved.
The Gaussian integral over the commuting $\phi$--$\bar\phi$
set of fields in \eGFActionI\ yields the determinant $\det^{-1}
\triangle_{(0)}$, which cancels with the $\det\triangle_{(0)}$
contribution coming from the anti-commuting set of fields $c$--$
\bar c$.

Now consider the term $\sqrt g C^2$. One could in principle work
out this term to second order in the metric variation $\delta
g_{ab}=h_{ab}$, but in practice this is an extremely tedious
exercise. The trick is to rewrite $\sqrt g C^2=\sqrt gC\ast C
+2\sqrt gW_-^2$, and realise that since the first term is a
topological invariant, it does not contribute to the quantum
dynamics. The second term, of the form $\sqrt gW_-^2$, is
positive definite and is minimised when the manifold is self-dual.
To second order in the trace-free metric variation $h_{ab}$, it is
\eqn\eWtoSecondOrder{\eqalign{&\sqrt g(1 - \hbox{$1\over4$}
h^2)(W_-+\D_1h)^2 \cr &\qquad= \sqrt g(W_-^2
+ (\D_1h)^2 + 2W_-\D_1h - \hbox{$1\over4$}h^2W_-^2)\ .}}
But imposing the condition $W_-=0$, we find that the second order
quantum variation coming from $C^2$ term is
\eqn\eWtoSecondOrder{(\D_1h)^2 \sim -h\D_1^\ast\D_1 h\ .}
Observe that the $h$ fields in \eWtoSecondOrder\ are transverse
trace-free. By contrast, the $h$ fields in the second term of
\eGFActionI\ are longitudinal. With both terms together, the
Gaussian integral over the $h$ field just gives $\det^{-1/2}
\triangle_{(1)}$ in the quadratic partition function.

Consider now the two remaining terms of \eGFActionI:
\eqn\eRemaningTerms{\bar\chi\D_1\psi + \bar\eta\D^\ast_0\psi\ .}
These $\bar\eta$--$\psi$--$\bar\chi$ fields are precisely
those encountered in the deformation complex \eSingerCplxII.
Observe that the $\bar\chi$ and $\bar\eta$ field equations from
\eRemaningTerms\ respectively give
\eqn\eFieldEqns{\D_1\psi=0\ ,\qquad \D^\ast_0\psi=0\ .}
These equations are nothing but the deformation equations.
Hence if we want to eliminate fermionic\foot{In this paper, the
word ``fermionic'' refers to the anti-commuting ghost fields,
but they do not have half-integer spins.} zero-modes from this
theory, we will have to choose our manifold $M$ such that its
gravitational moduli space has vanishing dimension. This is what
we will assume in the rest of this section.

A subtlety arises when trying to evaluate determinants for the
$\bar\eta$--$\psi$--$\bar\chi$ system. Note that the terms in
\eRemaningTerms\ define a differential operator T \nref\rBRT
{D.~Birmingham, M.~Rakowski and G.~Thompson, Nucl.~Phys. B329
(1990) 83}\refs{\rBRT,\rBBRT}
\eqn\eT{\T:\Omega_{22}\rightarrow\Omega_{11}\oplus\Omega_{40}\ ,}
which is the same operator as that defined in \eSingerCplxIII.
Since T does not map $\Omega_{22}$ into itself, its determinant
is not {\it a priori} well-defined. One could however consider
the adjoint of T
\eqn\eTStar{\T^\ast:\Omega_{11}\oplus\Omega_{40}\rightarrow
\Omega_{22}\ ,}
and define in the usual way
\eqn\eDetTi{\det\T\equiv\det{}^{1/2}(\T^\ast\T)\ ,}
where $\T^\ast\T$ maps $\Omega_{22}$ into itself. Thus, in this
case $\det\T=\det^{1/2}\triangle_{(1)}$. Hence, our quadratic
partition function reduces to simply
\eqn\eZQuadI{Z^{(2)}=\left({\det\triangle_{(1)}\over
\det\triangle_{(1)}}\right)^{1\over2}=\pm1\ .}
The total partition function is then a sum of $\pm1$'s over
conformally self-dual gravitational instantons on $M$:
\eqn\eZi{Z(M)=\sum_{\rm instantons}\pm1\ .}

We could have started out by choosing instead the delta-function
gauge: $\alpha=\beta=0$. This gauge has the advantage over the
previous choice of gauge in being computationally simpler,
although it does not illustrate the physical interpretation
of topological gravity as a type of conformal gravity theory
without local degrees of freedom. But the final results are the
same, as we now quickly show.

In the delta-function gauge, the quadratic partition function
in \eZQuad\ is
\eqnn\eQuadi
$$\eqalignno{I^{(2)}&=-\int_M \hbox{d}^4x\sqrt g\ \s\,\{
\bar\chi W_- + \bar c\D^\ast_0h + \bar\phi\D^\ast_0\psi\} \cr
&=-\int_M \hbox{d}^4x\sqrt g\ (BW_- + \bar\chi\D_1\psi
+ \bar\chi\D_1\D_0c + b\D_0^\ast h + \bar c\D^\ast_0\psi \cr
&\hskip1.00in + \bar c\D^\ast_0\D_0c + \bar\eta\D^\ast_0\psi
+ \bar\phi\D^\ast_0\D_0\phi)\ . &\eQuadi}$$
Again the determinants coming from the $\phi$--$\bar\phi$
term cancels with that of the $c$--$\bar c$ term, and we
absorb the $\bar c\D^\ast_0\psi$ term into the $\bar\eta
\D^\ast_0\psi$ term.

The functional integral over the $B$ field yields the delta
function constraint $\delta(W_-)=0$. This enforces the on-shell
condition automatically, so that $\D_1\D_0=0$. Recalling that
the lowest order variation of $W_-$ is $\D_1h$, the quadratic
partition becomes
\eqn\eQuadii{I^{(2)}=-\int_M \hbox{d}^4x\sqrt g\ (B\D_1h +
b\D_0^\ast h + \bar\chi\D_1\psi + \bar\eta\D^\ast_0\psi)\ .}
That this gives \eZQuad\ value one up to a sign is unambiguous.
The $\bar\eta$--$\psi$--$\bar\chi$ system of anti-commuting
fields, by previous arguments, yields the determinant term
$\det^{1/2}\triangle_{(1)}$. Analogously, the $b$--$h$--$B$
system of commuting fields gives $\det^{-1/2}\triangle_{(1)}$,
which cancels with the other determinant. Hence we arrive
at \eZi\ as before.

Note that this value of the partition function is conformally
invariant, so our quantum theory of topological gravity is
free of any conformal anomaly.

The expression \eZi\ is our gravitational analogue of the first
Donaldson invariant, which is the sum of $\pm1$'s over discrete
points of the Yang--Mills moduli space. To determine which sign
to use for each instanton, it would not be unreasonable to
proceed in the same way as in the Yang--Mills case \rTQFT. We
choose a particular instanton metric $g_0$ and declare it to
have the positive sign. Given any other instanton $g_1$, we
interpolate between the two via a curve in the space of conformally
and physically inequivalent metrics, and change the sign whenever
the fermionic determinant in \eZi\ has a zero-eigenvalue at some
point along the curve. Spectral flow guarantees that this
definition is unambiguous, i.e., it is independent of the curve
chosen.

It is clear that \eZi\ does not depend on the metric, since
we are in effect ``summing over metrics'' with a particular
topology. (By contrast, the metric independence of the
corresponding Donaldson invariant is less obvious \rDonaldson.)
For now, we will {\it conjecture} that \eZi\ is a differential
invariant of conformally self-dual four-manifolds with a
discrete gravitational moduli space. In other words, it is able
to distinguish inequivalent smooth structures underlying the
manifold, much like the Donaldson invariant. Because of the
heuristic nature of our quantum field theoretic derivation,
it would be very satisfying if a more rigorous derivation and
study of \eZi\ could be made.

In the case when the dimension of the moduli space is non-zero,
there will be non-trivial solutions to \eFieldEqns, resulting
in fermionic zero-modes in the theory. The general strategy
we then need to adopt is to introduce non-vanishing path
integrals of the form
\eqn\eCorrFunc{Z({\cal O})=\int{\cal D}X\ \exp(-I)
\cdot{\cal O}\ ,}
where ${\cal O}$ is some functional of the fields $X$. In the
field theoretic sense, ${\cal O}$ is called an observable and
$Z({\cal O})$ is the correlation function or vacuum expectation
value of the observable. In order to preserve the topological
nature of these correlation functions, we require ${\cal O}$ to
be s-invariant. We then absorb the zero-modes by demanding that
${\cal O}$ has ghost number equal to the dimension of the moduli
space \rTQFT.

\newsec{Concluding remarks}

In this paper, we have written down an elliptic complex
\eSingerCplxI\ which describes deformations of conformally
self-dual gravitational instantons. By applying the Atiyah--Singer
index theorem to this elliptic complex, we derived an expression
\eHI\ for the number of independent non-trivial deformations
that can be made about a given gravitational instanton, which
preserve its self-duality.

For a given manifold $M$, we defined the gravitational moduli
space of $M$ to be the set of conformally self-dual metrics on
$M$, factored out by conformal and coordinate transformations.
The virtual dimension of the moduli space is given by the number
\eHI.

Armed with this mathematical theory of instanton deformations,
we then proceeded to develop a theory of topological conformal
gravity starting from the classical action \eClassicalTG. The
BRST gauge-fixing procedure introduced three ghost fields
which characterise quantum fluctuations about classical
gravitational instantons. It turned out that these ghost fields
are most naturally described by the above elliptic complex.
In particular, when the gravitational moduli space consists
of discrete points, we evaluated the partition function of
topological gravity to obtain \eZi. This quantity may be regarded
as the gravitational counterpart to the first Donaldson invariant
of four-manifolds.

It may be possible to compute gravitational analogues of the
higher-order Donaldson invariants when the dimension of the moduli
space is greater than zero, by choosing appropriate observables
${\cal O}$ in \eCorrFunc. However, it would probably be premature
to do so here, at least until a proper mathematical theory
utilising gravitational instantons to classify four-manifolds
materialises. With such a theory, it would then provide an impetus
for physicists to try to rederive the invariants from topological
gravity. What we hoped was achieved in this paper, was to set up
the basic quantum theory of topological conformal gravity, and thus
put forward plausible physical evidence that gravitational
instantons, like their Yang--Mills counterparts, may be used
to study the differential topology of four-manifolds. To this
effect, we calculated the value of the partition function in a
simple case to give a flavour of how these invariants might look
like.

In a separate light, one may conjecture that a theory of quantum
gravity, if it exists, could be described by a topological
quantum field theory. What we live in may then correspond to
a phase of quantum gravity where the topological symmetry has
been broken, possibly in a way not unlike the Higgs mechanism.
To study such a scenario, we need to develop models of
topological gravity in four dimensions. One such model has
been proposed in this paper, and it is closely related to
conformal gravity.

There is another way in which four-dimensional topological
quantum field theories may be relevant to quantum gravity.
In the Euclidean path integral approach \ref\rEuclid{G.W.~Gibbons
and S.W.~Hawking eds., Euclidean Quantum Gravity (World
Scientific, Singapore, 1992)}, one is supposed to sum over all
metrics and topologies in the path integral. However, the
complete classification of four-manifolds still eludes us, so
any sum over topologies necessarily cannot be complete. This
does not mean that we should look for the key only under the
lamp-post where there is light, to quote Hawking \ref\rHawk{
S.W.~Hawking, Nucl.~Phys. B144 (1978) 349}. A way to get round
this is to postulate that any manifold invariant admits a path
integral representation via a topological field theory.
Consequently, we could possibly define a sensible measure
for the sum over topologies by considering topological
Yang--Mills theory, topological gravity and other
four-dimensional topological quantum field theories, rather
than working with the manifold invariants themselves.

Only time will tell.

\bigbreak\bigskip\bigskip\centerline{{\bf Acknowledgements}}
\nobreak\noindent
The results of this paper were presented by E.T.~at the
Les Houches Summer School ``Gravitation and Quantisations''
in July, 1992. He wishes to thank the participants for their
kind interest in this work, and the organisers for a very
stimulating School. He also gratefully acknowledges the receipt
of a Rouse Ball Travelling Studentship from Trinity College,
Cambridge, which financed the trip to Les Houches. M.J.P.~thanks
The Royal Society and Trinity College for financial support,
and Is Singer for a useful conversation.

\listrefs
\bye